\documentstyle[aaspp4]{article}     
\begin{document}
\slugcomment{Scheduled for ApJ, v536n1, Jun 10, 2000}
\title{Dithering Strategies for Efficient Self-Calibration of Imaging Arrays}
\author{Richard G. Arendt\altaffilmark{1}, D. J. Fixsen\altaffilmark{1},
\& S. Harvey Moseley}
\affil{Laboratory for Astronomy and Solar Physics,\\
Code 685, NASA GSFC, Greenbelt, MD 20771;\\
arendt, fixsen, moseley@stars.gsfc.nasa.gov}
\altaffiltext{1}{Raytheon ITSS}

\begin{abstract}
With high sensitivity imaging arrays,
accurate calibration is essential to achieve the limits of
detection of space observatories. One can simultaneously
extract information about the scene being observed and the
calibration properties of the detector and imaging system
from redundant dithered images of a scene.
There are large differences in the
effectiveness of dithering strategies for
allowing the separation of detector properties from sky
brightness measurements. In this paper, we quantify these
differences by developing a figure of merit (FOM) for dithering 
procedures based on their usefulness for allowing calibration on 
all spatial scales. 
The figure of merit measures how well the 
gain characteristics of the detector are encoded in the 
measurements, and is independent of the 
techniques used to analyze the data.
Patterns similar to the antenna arrangements 
of radio interferometers with good $u-v$ plane coverage, are found
to have good figures of merit. We present
patterns for both deep surveys of limited sky areas and for
shallow surveys. 
By choosing a strategy that encodes the calibration in the
observations in an easily extractable way, we enhance our
ability to calibrate our detector systems and to reach the
ultimate limits of sensitivity which are required to
achieve the promise of many missions.
\end{abstract}

\section{Introduction}
In order to achieve their required performance, many
observing systems must observe with sensitivities near
their confusion limits. 
Many instruments are capable of reaching these limits in crowded 
stellar fields such as the Galactic center. Future instruments
such as the Infrared Array Camera (IRAC) and the 
Multiband Imaging Photometer (MIPS) on the
{\it Space Infrared Telescope Facility} ({\it SIRTF}) and those
planned for the Next Generation Space Telescope (NGST) will
be able to reach limits in which the confusion of extragalactic
sources becomes significant.
In general, the measurement noise is determined by both the statistical
fluctuations of the photon flux and
uncertainties in detector gain and offset. Any successful
calibration procedure must determine these detector
parameters sufficiently accurately so that their uncertainties
make small contributions to the measurements errors
compared to those of the background fluctuations.
If the science done with the instrument requires
substantial spatial or temporal modeling,
calibration requirements become more demanding, ultimately
requiring similar integration time for observation and
calibration as in the case of the {\it COBE} FIRAS instrument
(Mather et al. 1994; Fixsen et al. 1994). Additionally, in such cases
robust error estimators are often needed. A common
method to determine the instrument calibration is to look at known
calibration scenes (e.g. a dark shutter, an illuminated screen, 
or a blank region of sky) 
of different brightnesses to deduce gain
and offset of each detector pixel. This requires a well
characterized calibration source and often a change in instrument
mode to carry out the measurement.
This procedure may introduce systematic errors relating to the
extrapolations from the time and conditions of the calibration observations
to the time and conditions of the sky observations and 
from the intensity (and assumed
flatness) of the calibration source to the intensity of the observed
sky. A different approach is to use the measurements of the sky alone to
extract the calibration data for the system. 
By using the sky observations for calibration, the systematic
errors introduced by applying a calibration derived from a distinctly
different data set are eliminated. Such methods
require a set of dithered images, where a single sky location is
imaged on many different detector pixels. 

Typical CCD and IR array data reduction procedures for a set of dithered 
images make use of a known or measured dark frame
($F^p$) and derive the flat field ($G^p$) through taking the weighted average
or median value of all data ($D^i$) observed by each detector pixel $p$ 
($i \in p$) in a stack of dithered images (e.g.
Tyson 1986, Tyson \& Seitzer 1988, Joyce 1992, Gardner 1995). 
The least squares solution of 
\begin{equation}
{\cal D}^i = G^p S^0
\end{equation}
where ${\cal D}^i = D^i - F^p$ ($i \in p$) and $S^0$ is the perfectly flat 
sky intensity, for $G^p$, the flat field, is
\begin{equation}
G^p = \frac{\sum_{i \in p}{\cal D}^i W_i}{\sum_{i \in p}W_i} \frac{1}{S^0}
\end{equation}
which is simply the weighted average of the data collected by each detector 
pixel normalized by the constant sky intensity (to be determined later though
the absolute calibration of the data). 
The weights, $W_i$, are normally determined by the inverse variance of the
data, but may also be set to zero to exclude sources above the background 
level.
The use of the median, instead 
of the weighted average, also rejects the outliers arising from 
the observations of real sources instead of the flat background, $S^0$, 
and formally corresponds to a minimization of the mean absolute deviation
rather than a least squares procedure. 
In either form, this method requires 
observations of relatively empty fields where variations in the background 
sky level are not larger than the faintest signal that is sought. 
Thus, throughout this paper we refer to such procedures as
``flat sky'' techniques.
As instrumentation improves and telescope sensitivity increases, 
this condition is becoming harder to fulfill. 
In fields at low Galactic latitude, stellar and nebular confusion 
can be unavoidable, and at high latitude deep imaging (particularly in 
the infrared) is expected to reach the extragalactic confusion limit.
In such cases, because of the complex background, and in other 
cases where external influences (e.g. moonlight, zodiacal light) create
a sky background with a gradient, the flat sky
approach does not work and a more comprehensive approach must be used.

Such an approach has been presented by Fixsen et al. (2000) who 
describe the general least squares solution for deriving 
the sky intensity $S^{\alpha}$ at each pixel $\alpha$, in addition to the 
detector gain (or flat field) $G^p$ and offset (or dark current + bias) $F^p$
at each detector pixel $p$, where each measurement, $D^i$, is represented by
\begin{equation}
D^i = G^p S^{\alpha} + F^p.
\end{equation}
(Throughout this paper we refer to the procedure described by Fixsen et al. 
(2000) as the ``least squares'' procedure.)
They show how the problem of inverting large matrices can be 
circumvented, and how the formulation of the problem allows for explicit 
tracking of the uncertainties and correlations in the derived $G^p$, $F^p$, 
and $S^{\alpha}$. Fixsen et al. also show that although the formal size of the 
matrices used in the least squares solution increases as $P^2$, where $P$ is 
the number of pixels in the detector array, the number of non-zero elements 
in these matrices increases only as $M\times P$, where $M$ is the number of 
images in the data set. In practice, the portion of the least squares solution 
for the detector gains and offsets is calculated first, and then the data 
are corrected to produce images of the sky ($S^{\alpha}$) that are 
registered and mapped into a final single image.
Because this approach explicitly assumes a different sky
intensity at each pixel, the crowded or confused fields that can cause the 
flat sky technique to fail are an aid to finding the least
squares solution. Thus, the need for chopping away from a complex source 
in order to observe a blank sky region is eliminated.
The simultaneous solution for both the detector gain and offset also 
eliminates the need for dark frame measurements, although if dark frame
measurements are available then they can be used with the other data to 
reduce the uncertainty of the procedure. 
We note that this general least squares approach may also be applied in 
non-astronomical situations (e.g. terrestrial observing) where complex images
are the norm.

The flat sky technique works well in situations where 
all detector pixels 
spend most of the time observing the same celestial calibration source,
namely the flat sky background. For this technique, dithering is required 
only to ensure that all pixels usually do see the background.
Because all pixels have observed the same
source, the relative calibrations of any two pixels in the detector are
tightly constrained, regardless of the separation between the pixels, i.e.
\begin{equation}
\frac{G^1}{G^2} = \frac{G^1 S^0}{G^2 S^0} = \frac{{\cal D}^1}{{\cal D}^2}.
\end{equation}
However, in the more general least squares solution of Fixsen et al. (2000), 
each sky pixel ($S^{\alpha}$) 
represents a different celestial calibration source. The only pixels
for which the relative calibrations are tightly constrained are 
those that through dithering have observed common sky pixels. Pixels 
that do not observe a common sky pixel are still constrained, though 
less directly, by intermediate detector pixels that do observe 
common sky pixels. For example, the relative calibration of detector
pixels 1 and 3 which observe sky pixels $\alpha$ and $\beta$ respectively,
but no common sky pixels, may be established if an intermediate detector
pixel 2 does observe both sky pixels $\alpha$ and $\beta$, i.e.
\begin{equation}
\frac{G^1}{G^3} = 
\frac{G^1 S^{\alpha}}{G^2 S^{\alpha}} \frac{G^2 S^{\beta}}{G^3 S^{\beta}} = 
\frac{{\cal D}^{1\alpha}}{{\cal D}^{2\alpha}} \frac{{\cal D}^{2\beta}}{{\cal D}^{3\beta}}.
\end{equation}
Other detector pixels might require multiple intermediate pixels to 
establish a relative calibrations. As the chain of intermediate
pixels grows longer, the uncertainty of the relative calibration of the
two pixels also grows. Therefore, when applying the least squares solution, 
the exact dither pattern becomes much more important than in the 
flat sky technique. For the least squares solution to 
produce the smallest uncertainty, 
the dither pattern should be one that establishes the tightest 
correlations between all pairs of detector pixels using a small
number of dithered images. Even if one is only interested 
in small scale structure on the sky (e.g. point sources), it is still 
important to have the detector properly calibrated on {\it all} spatial 
scales to prevent large scale detector variations from biasing results 
derived for both sources and backgrounds imaged in different parts of 
the array.

Whether obtained by flat sky, least squares, or other techniques, 
the quality of the calibration is ultimately determined by its
uncertainties. For the least squares solution of Fixsen et al. (2000), 
understanding the uncertainties is relatively straight forward, because it
is a linear process, i.e. $P^{\alpha} = L^{\alpha}_i D^i$ where $P^{\alpha}$
is the set of fitted parameters, $D^i$ is the data, and $L^{\alpha}_i$ is 
a linear operator. Then, given a covariance matrix of the data, $\Sigma^{ij}$,
the solution covariance matrix is 
$V^{\alpha\beta} = L^{\alpha}_i L^{\beta}_j \Sigma^{ij}$. 
For a nonlinear process such as a median filter the uncertainties are 
harder to calculate. The diagonal terms of the covariance matrix of the 
solution might be
sufficiently well approximated by Monte Carlo methods, but the off-diagonal
components are far more numerous and often more pernicious as the effects
can be more subtle than the simple uncertainty implied by the diagonal
components. For this reason the off-diagonal components are often ignored. 
Creating final images at subpixel resolution (e.g. ``drizzle'', Fruchter
\& Hook 1998) may introduce additional correlations beyond those
described by the covariance matrix, and disproportionately 
increase the effects of the off-diagonal elements of the correlation matrix. 
Accurate knowledge of all these 
uncertainties is especially important for studies that seek spatial 
correlations within large samples, such as deep galaxy surveys
or studies of cosmic backgrounds, so that any detected correlations are 
certifiably real and not artifacts caused by the calibration errors
and unrecognized because of incomplete or faulty knowledge of the 
uncertainties.

Table 1 itemizes some of the features of each data analysis technique.
The remainder of this paper is concerned with characterizing what makes a
dither pattern good for self-calibration purposes using the least squares 
solution. We present a ``figure of merit'' (FOM) which can be used as a
quantitative means of ranking
the suitability of different dither patterns (\S2). We then present
several examples of good, fair and poor dither patterns (\S3), and investigate
how changes to the patterns affect their FOM. In \S4, we show how dithered
data can be collected in the context of both
deep and shallow surveys. We also
investigate the combined effects of dithering and the survey grid geometry on
the completeness of coverage provided by the survey. Section 5 discusses
miscellaneous details of the application and implementation of dithering.
Section 6 summarizes the results.

\section{Evaluation of Dithering Strategies}
\subsection{Dithering}
To be specific, we define the process of ``dithering'' as obtaining
multiple mostly overlapping images of a single field. 
Normally, each of the dithered 
images has a different spatial offset from the center of the field, and none
of the offsets of the dither pattern is larger than about half of the size 
of the detector array.
Generally, the set of dithered images is averaged in some manner into a 
single high-quality image for scientific analysis.
This is distinct from the processes of ``surveying'' or ``mapping'', in which
a field much larger than the size of the array is observed, using images that
are only partially overlapping. If survey data is combined into a 
single image for analysis, then the process required is one of 
mosaicking more than averaging.
A region may be surveyed or mapped using dithered images at each of the survey
grid points. 

There are several reasons why an observer might wish to collect dithered data.
One is simply to make sure that no point in the 
field remains unobserved 
because it happened to be targeted by a defective pixel in the detector array.
To meet this objective, two dither images would suffice, provided their 
offsets are selected to prevent two different bad pixels from targeting the 
same sky location. A second reason to dither is so that point
sources sample many different subpixel locations or phases. Such a data set 
allows recovery of higher resolution in the event that the detector pixel 
scale undersamples the instrumental point spread function. Several procedures
have been developed for this type of analysis, which is commonly applied to 
HST imaging data and 2MASS data (e.g. Fruchter \& Hook 1998; Williams 
et al. 1996; Lauer 1999; Cutri et al. 1999). 
A third reason to dither is to obtain a data set which contains sufficient 
information to derive the detector calibration and the sky intensities from 
the dithered data alone. As discussed in the introduction, 
for the flat sky approach, the flatness of the background is a 
more important concern than the particular dither pattern. However, 
this is reversed when the least squares solution to the calibration is 
derived (Fixsen et al. 2000). The structure of the sky is less important 
than the dither pattern which needs to be chosen carefully so that the 
solution is well-constrained. 

In an attempt to cover as wide a field as possible, the detector array
often undersamples the instrument point spread function. This undersampling 
can lead to increased noise in the least squares calibration procedure. 
There are several ways this extra noise can be alleviated. One way is
to use strictly integer-pixel offsets in the dither pattern. However, 
this requires very precise instrument control, and eliminates the possibility 
of reconstruction of the image at subpixel resolution (i.e. resolution 
closer to that of the point spread function). A second way to reduce noise is
to assign lower weights to data where steep intensity gradients are present.
A third way of dealing with the effects of undersampled data is to use
subpixel interlacing of the sky pixels within the least squares solution 
procedure. This technique may require additional dithering over the region 
since the interlaced sky subpixels are covered less densely than full 
size pixels. A fourth means is that the least squares procedure of 
Fixsen et al. (2000) could be modified to account for each datum ($D^i$) 
arising from a combination of several pixel (or subpixel) sky intensities 
($S^{\alpha}$). This is a significant complication of the procedure. 

After the least squares method is used to derived the detector 
calibration, users can always apply the method of their choice (e.g. 
``drizzle'' described by Fruchter \& Hook 1998) 
for mapping the set of calibrated images into a single subpixelized image. 
Such methods may or may not allow continued tracking of the uncertainties and
their correlations that the least squares procedure provides.

Dithering involves repointing the telescope or instrument, 
and thus may require additional
time compared to simply taking multiple exposures of the same field. 
Multiple exposures of the same field without dithering would 
allow rejection of data affected
by transient effects (e.g. cosmic rays), and improved sensitivity 
through averaging exposures, but of course lack the benefits described above.
Whether the time gained by not dithering outweighs the benefits lost, 
will depend on the instrument and the observer's scientific goals.

\subsection{A Figure of Merit}
The accuracy of the calibration of an array detector cannot be fully specified
by a single number or even a single number per detector pixel. The
full covariance matrix is necessary to provide a complete description 
of the uncertainties. The magnitude of the diagonal elements 
of the covariance matrix (i.e. $\sigma^2_p$) is determined primarily by
the noise characteristics of the instrument and the sky, and is sensitive
to the number of images collected in a set of dithered data, but not to
the dither pattern. The off-diagonal elements of the covariance matrix 
are sensitive to the dither pattern, and through the correlations they 
represent, any measurements made from the calibrated data will contain
some imprint of the dither pattern. 
(In general these correlations degrade the signal quality although they
can improve the results of some types of measurements depending on
whether the correlations are positive or negative and whether the two
data elements are used with the same or opposite sign in the measurement.)
In order to obtain the best calibration,
one would like to use a dither pattern that minimizes the correlations 
it leaves in the calibrated data. Since comparison of the entire 
covariance matrices for different dither patterns is awkward, we adopt
a single number, a ``figure of merit'', that is intended to provide a generic
measure of the relative size of the off diagonal terms of the covariance 
matrix. 
The figure of merit (FOM) is designed only to compare different dither 
patterns rather than investigating all of the details of a full 
observing system (i.e. particular telescope/instrument combinations). 
The instrumental details matter of course, and in practice they may 
place additional constraints in choosing the dither pattern. 

Here we make several simplifying assumptions to ease the 
calculations and comparisons. First we assume that all of the 
detector pixels have approximately the same noise and gain.
Next we assume that the noise is independent of sky position, either because
the Poisson counting statistics are not important or the observed field
is so uniform that the photon counting statistics do not vary appreciably
across the field.
With these assumptions we can simultaneously solve for both the 
gain and/or offset for each detector pixel and the sky brightness of 
each sky pixel (Fixsen et al. 2000). The
solution necessarily introduces correlations into the uncertainties.

For the figure of merit we choose only a single pixel at the center 
of the array and look at its correlations. This
is done to reduce the calculational burden which includes 4 billion 
correlations for a modest $256\times 256$ detector. Since all of the pixels are
locked to the same dither pattern the correlations are similar for the other
pixels (discussed below).
We sum the absolute value of the correlations between the 
central pixel and all of the other pixels. 
This is compared with the variance of the central pixel, 
$\sigma^2_{p_0}$,
as this is the irreducible uncertainty due to detector noise alone. 
Thus, we define the figure of merit ($FOM$) as:
\begin{equation}
FOM = \frac{\sigma^2_{p_0}}{\sum_{i\in {\rm\ all\ pixels}} |V_{ip_0}|}
\end{equation}
where $V$ is the covariance matrix of the detector parameters. 
The absolute value is used here to ensure that the sum will be small
only if all of the terms are small, not because some of the frequent
negative correlations happen to cancel the positive correlations.
In detail,
the FOM is a function ($f(x) \approx 1/(1+x)$) of the mean absolute value
of the normalized off-diagonal elements of the covariance matrix.
With this definition, the FOM is bounded on the range [0,1], and can 
be thought of as an efficiency of encoding correlations in the dither 
pattern, i.e. a high FOM is desired in a dither pattern. 

Equation 6 is not unique. A wide variety of possible quantitative
figures of merit could be calculated. Ideally one would choose the FOM that 
gives the lowest uncertainties in the final answer. This can be done if the 
question, i.e. quantity to be measured or scientific goal, 
is well determined. In that case the question can be posed as a vector
(or if there is a set of questions, a corresponding 
set of vectors in the form of a
matrix). The vector (or matrix) can then be dotted on either side of the
covariance matrix and the resulting uncertainty minimized. There are several
problems in this approach. One is that the matrix is too large to 
practically fit in most computers. A second problem is that the question may 
not be known before the data are collected. A third problem is that the same 
data may be used to answer several
questions. To deal with the first issue we use only a single row or column
of the symmetric covariance matrix. As shown below, the rows of the matrix
have a similar structure over most of the array. To deal with the other two
issues, the FOM uses the sum the absolute value of all of the terms. This may 
not be the ideal FOM for a specific measurement, but it should be a good
FOM for a wide variety of measurements to be made from the data. 

Throughout this paper, we calculate the FOM based on calibration 
which only seeks to determine the detector gains or offsets, but not both.
When both gains and offsets are sought, the solution for the covariance
matrix contains degeneracies that are only broken by the presence 
of a non-uniform sky brightness (Fixsen et al. 2000). The FOM when solving
for one detector parameter is similar to that which would 
apply when solving for both gains and offsets.

\subsection{Dither Patterns and Radio Interferometers}
In order to compute relative gain and/or offset, two detector pixels must
observe the same sky pixel or have a connection through other detector
pixels that mutually observe one or more sky pixels. A shorter path of 
intermediate detectors implies a tighter connection and lower uncertainties.
One goal of dithering is to tighten the connections between
detectors and thus lower the uncertainties. This combinatorial problem happens
to share geometrical similarities with another problem that has 
been dealt with previously,
namely covering the $u-v$ plane with a limited number of antennas in a radio 
interferometer. 

Figure 1 shows the $u-v$ coverage of the VLA for a snapshot of a source at 
the zenith. Each antenna pair leads to a single sample marked with a dot 
in the $u-v$ plane. Also shown is the map of $|V_{ip_0}|$ generated by using
a 27-position dither pattern with the same geometry as the 
VLA array (\S3.2). The strongest correlations are found at locations 
of the direct dither steps corresponding to the VLA baselines.
However, the non-zero correlations (and anti-correlations) found elsewhere in 
the map make a significant contribution to the total FOM.
\begin{figure}[t]
\plotone{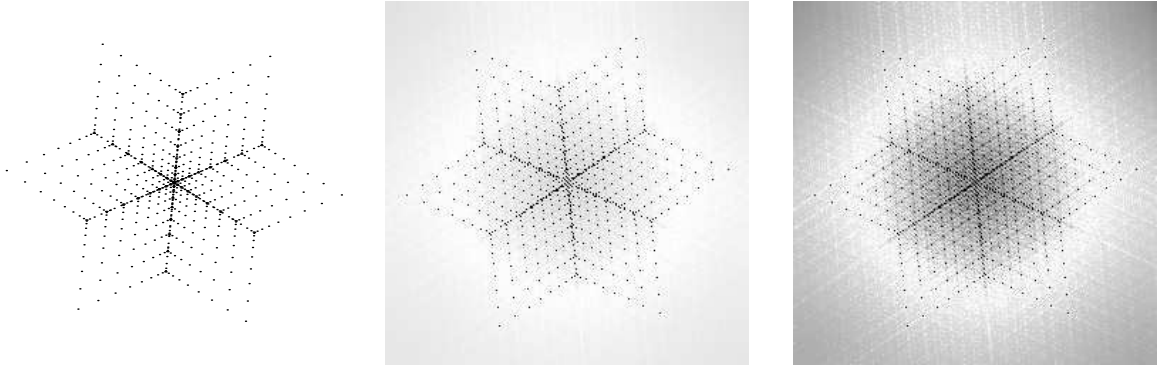}
\caption{On the left is the $u-v$ baseline coverage of the VLA for
a snapshot of a source at the zenith. In the center and on the right is
the map of $|V_{ip_0}|$ for a ``VLA'' dither pattern, stretched to emphasize
the similarity to the VLA $u-v$ plane coverage, and the weaker correlations
respectively.}
\end{figure}

Figure 2 shows maps of $|V_{ip_0}|$ generated using different choices of
$p_0$. These maps illustrate that the correlations of all pixels are similar 
in structure to those of the central pixel, but the finite size of the 
detector limits the correlations available to pixels near the detector edges.
The dither pattern used in this demonstration is the VLA pattern
described in \S3.2.
\begin{figure}[t]
\plotone{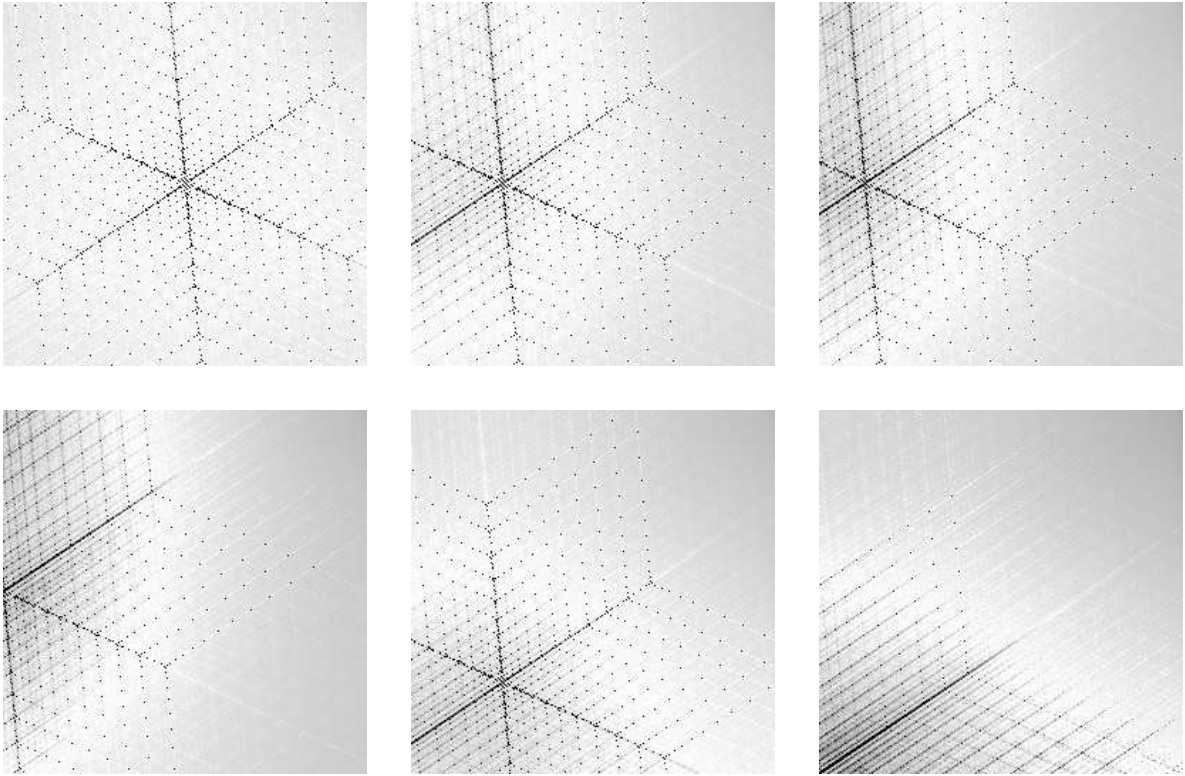}
\caption[]{The panels show the $|V_{ip_0}|$ correlations for detectors at 
the locations (128,128), (64,128), (32,128), (0,128), (64,64), 
and (0,0) in a $256\times256$ array 
(left to right and top to bottom). Dark spots represent 
strong correlations. The dither pattern used to calculate
these correlations is a 27-point VLA pattern.}
\end{figure}

Despite the similar geometries of radio interferometer $u-v$ coverage and 
dither pattern maps of $|V_{ip_0}|$, several important differences should be
noted. 
First, with radio telescopes only direct pairs of antennas
(although all pairs) can be used to generate interference patterns, whereas
with dither patterns a path involving several 
intermediate detector pixels can be 
used to generate an indirect correlation. However, the greater the number of 
intermediate steps that must be used to establish a correlation, the noisier 
it will be. 
Second, the $u-v$ coverage is derived instantly. Observing over a 
period of time fills in more of the $u-v$ plane as the earth's rotation 
changes the interferometer baselines relative to the target source. In 
contrast, the $|V_{ip_0}|$ coverage shown in Figs. 1 and 2 is only achieved
after collecting dozens of dithered images. To fill in additional coverage,
the dither pattern must be altered directly because there is no equivalent
of the earth rotation that alters the geometry of the instrument with 
respect to the sky. Another important difference is that the short
interferometer baselines (found near the center of the $u-v$ plane) are
sensitive to the large-scale emission. For dither patterns the inverse 
relation holds. Direct correlations between nearby detector pixels are 
sensitive to small-scale structure in the detector properties and sky
intensities. Thus the outer edge of the interferometer's $u-v$ coverage
represents a limit on the smallest-scale structure that can be resolved,
while the outer edge of strong $|V_{ip_0}|$ correlations represents a 
limit on the largest-scale variations that can be reliably distinguished.

Overall, the geometrical similarities suggest that patterns used and 
proposed for radio interferometers may prove to be a useful basis set for
constructing dither patterns. In the following section, we calculate the FOM
for several patterns inspired by radio interferometers in addition to 
other designs.

\section{Various Dither Patterns}
Several general algorithms for generating dither patterns have been examined. 
In many cases, we have also explored variants of the basic algorithms by
changing functional forms, adding random perturbations, or applying overall
scale factors. We have also tested several specific examples of dither patterns
from various sources. Examples of the patterns described below are shown in 
Figure 3. All tests reported here assumed detector dimensions 
of $256\times 256$ pixels unless otherwise noted.
\begin{figure}[t]
\plotone{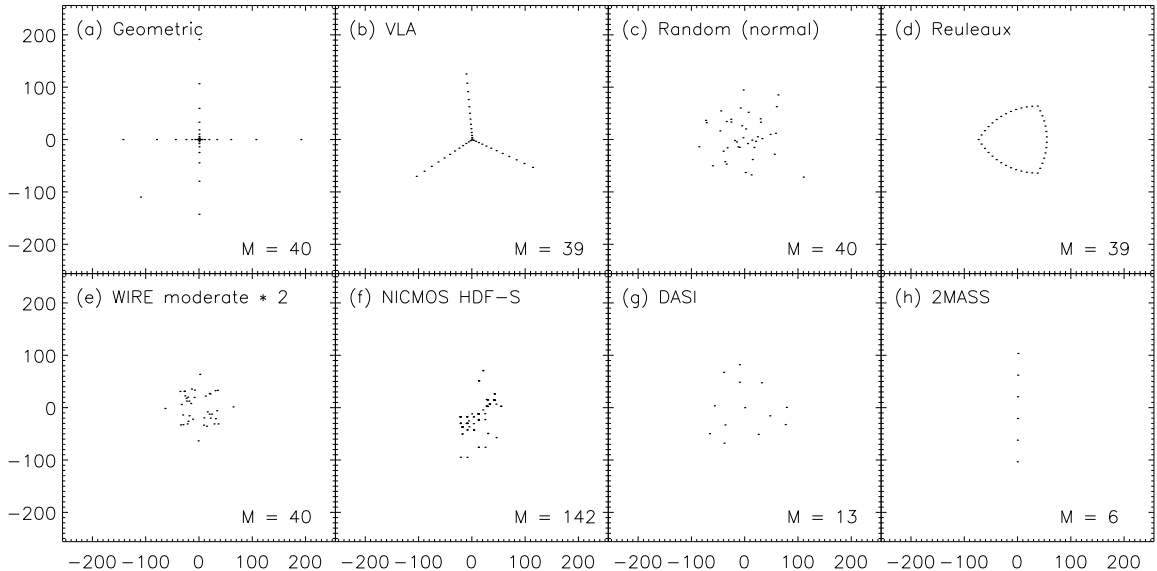}
\caption[]{Examples of some of the tested dither patterns. The dots
mark the center of the array for each of the $M$ positions for each pattern.}
\end{figure}

\subsection{Reuleaux Triangle}
Take an equilateral triangle and draw three 60\arcdeg\ arcs connecting each pair
of vertexes, while centered on the opposite vertex. The resulting fat triangle 
is a Reuleaux triangle. This basic shape has been used to set the geometry of
the Sub-Millimeter Array (SMA) on Mauna Kea (Keto 1997). 

This shape can be used as a dither pattern by taking equally spaced steps
along each side of the Reuleaux triangle. The length of the steps is set
by the overall size of the triangle (a free parameter) and the number of
frames to be used in the pattern. For an interferometer, Keto (1997) shows 
that the $u-v$ coverage can be improved by displacing the antennas from 
their equally spaced positions around the triangle. 

\subsection{VLA}
The ``Y''-shaped array configurations of the Very Large Array (VLA) radio 
interferometer are designed such that the antenna positions from the center
of the array are proportional to $i^{1.716}$ (Thompson, et al. 1980).
The three arms of the array are separated from each other by 
$\sim 120$\arcdeg. We have adopted this geometry to provide a dither pattern
with positions chosen along each of the three arms at 
\begin{equation}
dr = \sqrt{dx^2+dy^2} = i^p {\rm \ \ where\ \ } i = 1,2,3,...,M/3.
\end{equation}
\noindent and $p$ is an arbitrary power which can be used to scale 
the overall size 
of the pattern. The first step along each of the 3 arms is always at 
$dr = 1.0$. The azimuths of the arms were chosen to match those of the VLA, 
at 355\arcdeg, 115\arcdeg, and 236\arcdeg. 

\subsection{Random}
Random dither patterns were tested using $dx$ and $dy$ steps generated 
independently from
normal (Gaussian) or from uniform (flat) distributions. The widths of the 
normal distribution or the symmetric minimum and maximum of the uniform 
distribution are free parameters. 

\subsection{Geometric Progression}
We have generated a geometric progression pattern, stepping in $x$ in 
steps of $(-f)^n$, where
$n=0,1,...N-1$ and $f^N=256$. The same steps are also used in the $y$ direction.
This pattern separates the $x$ and $y$ dimensions. 
In each dimension the pattern is
quite economical in generating correlations up to the point where $f=2$. 
Beyond this
there is little to be gained in adding more dither steps in the $x$ or $y$
direction. However, there is some benefit expected in adding steps combining
$x$ and $y$ offsets. Hence, for a $256\times256$ array, we should expect the 
geometric pattern to be good for $M\leq 2\ log_2(256)=16$ positions 
and not show much improvement by adding more positions.

The geometric progression patterns used here contain two additional steps 
chosen at $(dx,dy) = (0,0)$ and at a position 
such that $\sum dx = \sum dy = 0.0$.
This is a cross-shaped pattern, with one diagonal pointing,
from which any desired pixel-to-pixel correlation can be made 
with a small number of intermediate steps. The alternating sign of the steps 
builds up longer separations quickly. 

\subsection{Other Patterns}
Several other patterns were also tested with little or no modifications. 
The patterns that were planned for the WIRE moderate and deep 
surveys were examined with both the nominal dither steps, and with steps
scaled by a factor of 2 to account for the difference between the 
$128\times128$ pixel WIRE detectors and a larger $256\times256$ pixel 
detector. The pattern used for NICMOS observations of the HDF-S was tested.
The configuration of the 13 antennas of the Degree Angular Scale 
Interferometry (DASI; Halverson, et al. 1998) was used as a scalable 
pattern. The declination scanning employed by 2MASS yields a linear dither 
pattern. 

\subsection{Figures of Merit for the Patterns}
In the simplest form, a specific pattern, $M$ images deep, would be used to 
collect data at a single target. The FOM for all patterns tested, with various
$M$ and other modifications, are listed in Table 2. 
For all patterns, the FOM increases (improves) as $M$ increases. For $M < 20$
the change is quite rapid. The variations of FOM as a function of $M$ for the 
tabulated versions of each of the patterns are shown in Figure 4. 
\begin{figure}[t]
\plotone{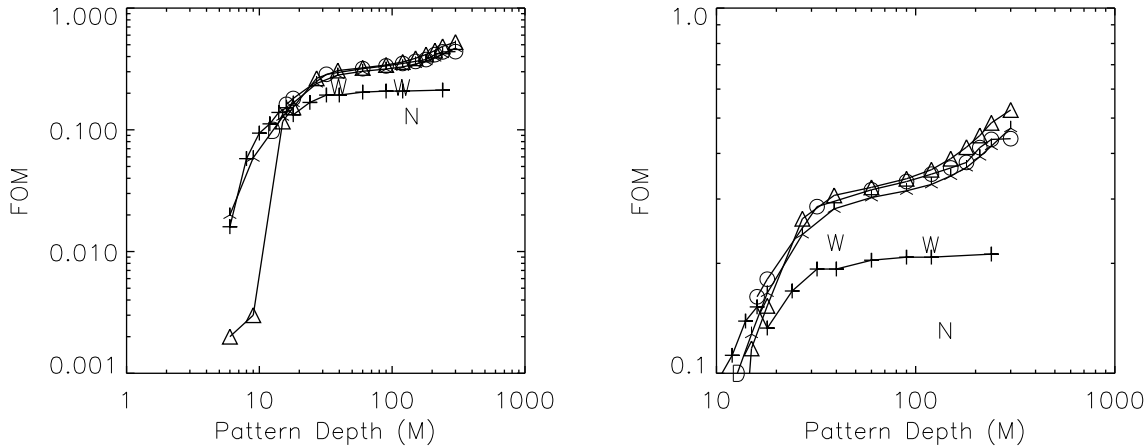}
\caption[]{The FOM as a function $M$ the number of positions in each pattern.
``+'' = geometric progression, inverted ``Y'' = VLA, 
``{\small $\bigcirc$}'' = random (normal), 
``$\bigtriangleup$'' = Reuleaux triangle, ``W'' = WIRE moderate and
deep surveys, ``D'' = DASI, and ``N'' = NICMOS coverage of the HDF-S.
The right panel shows the same data on an enlarged scale.}
\end{figure}

Table 2 also lists results for a Reuleaux triangle pattern applied to a 
$32 \times 32$ detector, and for two large grid dither patterns applied
to the same array. The grid dither patterns are square grids with 1 pixel 
spacings between dithers, such that for the $M = 1024$ pattern
a single sky pixel is observed with each detector, and for the $M = 4096$ 
pattern a $32 \times 32$ pixel region of sky is observed with each detector
pixel. These results demonstrate that in the extreme limit where all 
correlations are directly measured, the $FOM \rightarrow 1.0$. The FOM 
does not reach 1.0 because of the finite detector and dither pattern sizes.

For the $256 \times 256$ arrays, the Reuleaux and random 
(normal) patterns have the best FOM for $M > 20$.
The VLA pattern is only a little worse, but other patterns have distinctly 
smaller FOM than these patterns. For the scalable VLA, random, Reuleaux, and
DASI patterns, the best FOM for a fixed $M$ usually occurs when 
the maximum $|dx|$ or $|dy| \approx 128$ pixels. For patterns with 
small $M$ the optimum scale factor is usually smaller, to avoid too many large 
spacings between widely scattered dither positions.
For values of $M<20$ no pattern seems to produce a good FOM, however, 
the geometric pattern usually does best in this regime. 
Rotating the patterns with respect to the detector array generally produces
only modest changes in the FOM. For $M \lesssim 30$, 
the FOM of a Reuleaux pattern is improved by adding small random perturbations
to the dither positions. No optimization of the perturbations was performed
(as Keto 1997), but apparently any perturbation is better than none for small
$M$ patterns. Deep Reuleaux triangle patterns are neither improved nor 
worsened by small perturbations.

The results presented in Fig. 4 and Table 2 indicate that a good
FOM is dependent on patterns that sample a large number and 
wide range of spatial scales. 
A variety of patterns with different geometries can yield satisfactory 
results, as demonstrated by the rather different Reuleaux triangle
and random patterns. Therefore, attempts to find the single ``optimum'' 
pattern may not be very useful, and selection of a dither pattern needs
to carefully avoid patterns that contain obvious or hidden redundancies
that lead to a poor FOM. An example of this sort of pitfall is the $M = 18$
geometric pattern, for which all dither steps are integer powers of 2, 
leading to a FOM that is worse than geometric patterns with depths of $M = 14$ 
or 16. 

The coverage of the VLA, random, and Reuleaux triangle dither patterns when 
used for observation of a single target is shown as maps in Figure 5, and 
histograms in Figure 6. The Reuleaux triangle dither pattern provides the
largest region covered at maximum depth, but if a depth less than the 
maximum is still useful then the VLA dither pattern may provide the 
largest area covered.
\begin{figure}[t]
\plotone{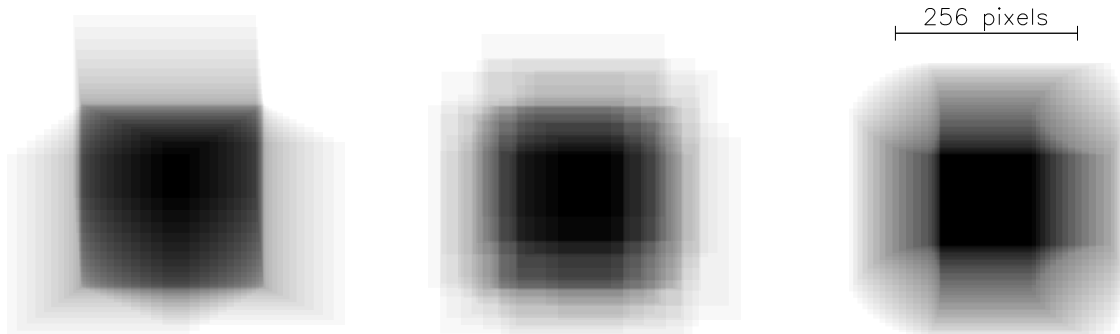}
\caption[]{Coverage maps for $M=39$ single target dither patterns
(left) VLA: $FOM = 0.282$, (center) Random Gaussian: $FOM = 0.302$, 
(right) Reuleaux triangle: $FOM = 0.307$.}
\end{figure}
\begin{figure}[t]
\plotone{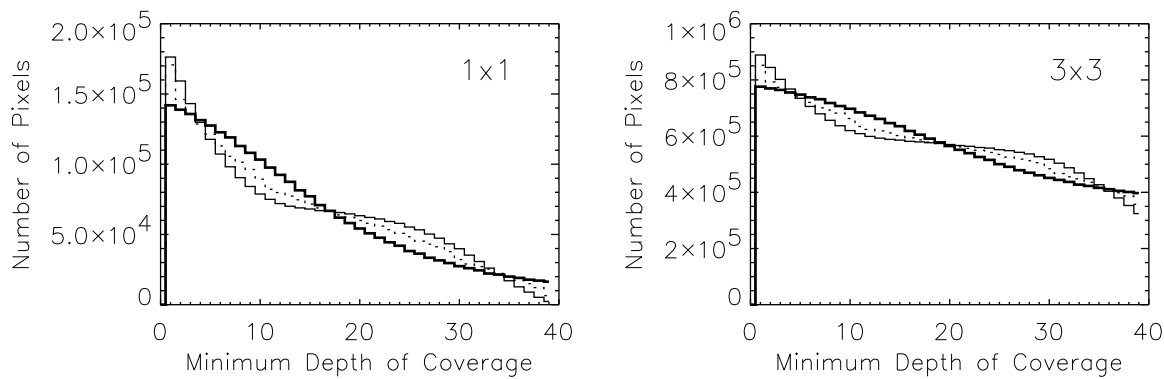}
\caption[]{Cumulative histograms of the coverage as a function
of minimum depth for $M=39$ VLA (thin), random Gaussian (dotted), and 
Reuleaux triangle (thick) dither patterns. Coverage for a single target is 
shown at left; coverage for a deep $3\times 3$ survey with a $256\times 256$ 
pixel grid spacing is shown at right.}
\end{figure}

The importance of the largest dither steps in a pattern is demonstrated 
through analysis of simulated WIRE data. A synthetic sky was sampled using 
both geometric progression and random dither patterns. The maximum dither
offset was 38 pixels for the geometric progression pattern and 17 pixels
for the random pattern. The FOM for this geometric pattern is 0.127, and for 
this random pattern it is 0.099. WIRE's detectors were $128\time 128$ arrays. 
The gain response map used in the simulations contained large scale 
gradients with amplitudes of $\sim 10\%$. Figure 7 shows comparisons
between the actual gains and the gains derived when the self-calibration 
procedure described by Fixsen, et al. (2000) is employed. 
The random dither pattern 
without the larger dither offsets was less effective at identifying the
large scale gain gradient. The undetected structure in the gain winds up 
appearing as a sky gradient that affects the photometry of both the 
point sources and the background in the images.
\begin{figure}[t]
\plotone{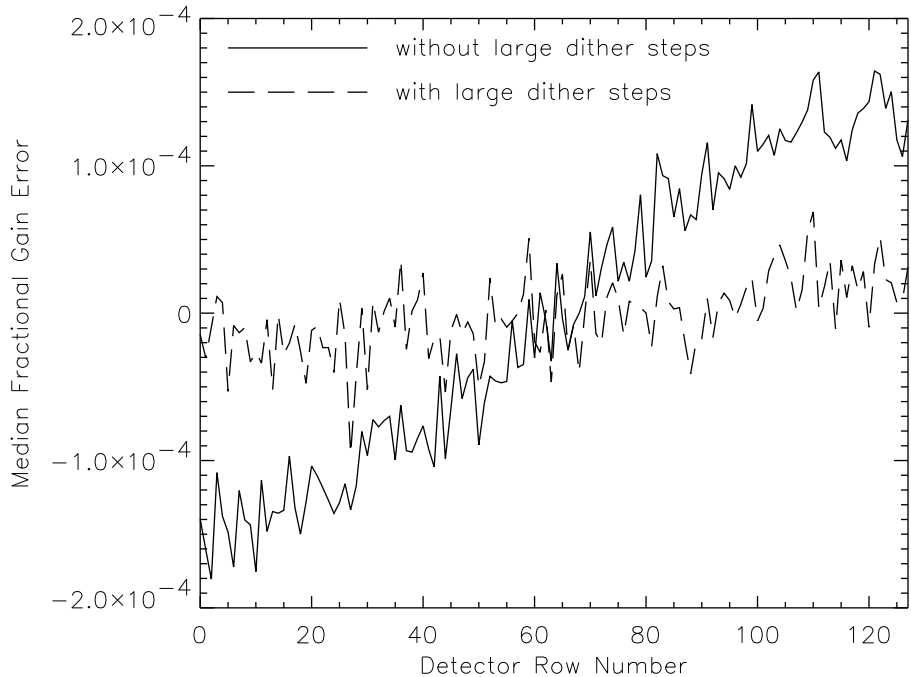}
\caption[]{The median fractional gain errors are plotted as a 
function of detector row
for detector gains derived from two simulated WIRE data sets. 
Each simulation contains 10 dithered images. Only one simulation includes 
relatively large dither steps. 
When applying a self-calibration algorithm, a lack of large dither steps 
leads to large-scale gain errors.}
\end{figure}

\section{Surveys}
\subsection{Deep Surveys}
For obtaining a standard deep survey, we have assumed that the same dither 
pattern is repeated at each location of a grid. The survey grid is assumed to 
be aligned with the detector array and square, with a spacing 
no larger than the size of the array. The FOM for surveys using several 
different dither patterns and grid spacings are listed in Table 3.
The FOM derived for the entire survey as a single data set is basically 
determined by the FOM of the dither pattern used. 
The overlap between dithers from
adjacent points in the survey grid, effectively adds additional 
steps to the dither pattern, which slightly improves the FOM over that of the 
pattern when used for a single target. Smaller survey grid spacings lead
to increased overlap and increased FOM, but also lead to a smaller area of
sky covered in a fixed number of frames. 
The improvement in the FOM when used in surveys rather than singly is most 
significant for relatively shallow dither patterns, however, even in a survey,
the FOM of a shallow pattern is still not very good. The FOM improves only
slightly as the survey grid grows larger than the basic $2\times 2$ unit cell.

\subsection{Shallow Surveys}
For shallow surveys in which as few as 2 images per grid location are 
desired, using 
the same small $M$ dither pattern at each location yields a very poor FOM.
An alternate method of performing a shallow survey is to choose a larger $M$
dither pattern and apply successive steps of the dither pattern at successive
locations in the survey grid (Figure 8). If the survey is large enough, 
it can contain 
all the direct correlations of the large $M$ dither pattern, though spread out
among many survey grid points rather than at a single location. The FOM of the 
shallow survey can thus approach the FOM of the single deeper dither pattern.
The advantage of altering the dither pattern at each survey grid point is 
still present, though less significant, as the survey depth increases. 
The FOM derived from various surveys using this shallow survey strategy are 
shown in Table 4. 
\begin{figure}[t]
\epsscale{0.5}
\plotone{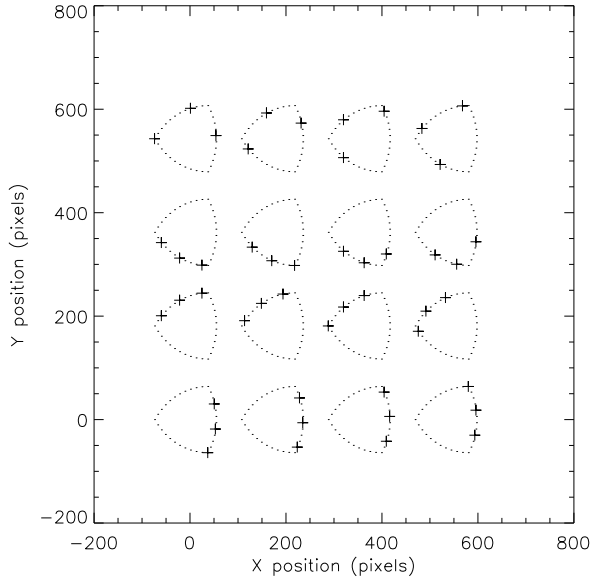}
\caption[]{An example of a $4\times 4$ $M=3$ shallow survey 
on a $181\times 181$ pixel grid using an $M=33$ Reuleaux triangle 
dither pattern. The dots show the repetition of the full dither pattern, 
while the crosses mark the dither points that were actually used at 
each survey grid point.}
\end{figure}

A random dither pattern is a natural choice for use in this shallow survey
strategy. One can proceed by simply generating a new random set of dithers
at each survey grid point. If a more structured dither pattern 
is used as the basis for the shallow survey (e.g. the Reuleaux 
triangle in Fig. 8), then one must address the
combinatorial problem of selecting the appropriate subsets of the larger 
dither pattern at each survey grid point. The example shown in Fig. 8 is
{\it not} an optimized solution to the combinatorial problem.

\subsection{Survey Coverage \& Grids}
When a large area is to be observed, the most efficient way to cover the 
region is to use a square survey grid aligned with the detector array and
with a grid spacing equal to the size of the array, or slightly less to
guard against bad edges or pointing errors. In this mode a deep survey 
using the same $M$ position dither pattern at each survey grid point
will cover the desired region at a depth of $M$ or greater. There will
be no holes in the coverage, though the edges of the surveyed region will
fade from coverage of $M$ to 0 with a profile determined by the dither 
pattern used (Fig. 6). A shallow survey, using a different dither pattern at
each grid point, may or may not have coverage holes depending on the 
maximum size of the dither steps and the grid spacing of the survey. The 
constraint for avoiding coverage holes is that the overlap of the 
survey grid must be more than the maximum range of dither step offsets 
(independently in the $x$ and $y$ coordinates), e.g.
\begin{equation}
X - \Delta X > {\rm max}(dx_i) - {\rm min}(dx_i)
\end{equation}
where $X$ is the size of the array, $\Delta X$ is the survey grid spacing, and
$dx_i$ are the dither steps ($i=1...M$). This constraint places the survey grid
points close enough together that coverage holes are avoided even if dithers
at adjacent grid point are offset in the maximum possible opposite directions. 
If the shallow survey observing program can be arranged to avoid this worst 
case, then the grid spacing may be increased without developing coverage holes.
Coverage holes may be undesirable when mapping 
an extended object, but may be irrelevant if one is simply seeking 
a random selection of point sources to count.
Note that some minor coverage holes are inevitable, where data are lost to 
bad pixels or cosmic rays. Additionally, a coverage hole where a depth of $M=1$
is achieved instead of $M=3$ might be more serious than one where $M=18$ is
achieved instead of $M=20$.

For this shallow survey strategy there is an inherent tradeoff between 
the area covered (without holes) and the FOM. Using a dither pattern 
containing large dither steps as the basis for the survey will lead to 
a good FOM, but require a relatively large overlap in the survey grid
spacing and a consequent loss of area covered by the survey. Decreasing
the scale of the dither pattern leads to a lower FOM, but permits an 
increase in the survey grid spacing and total area covered. The ideal 
balance between these will depend on the instrumental characteristics 
and the scientific objectives.

In many instances, an observer may want to survey or map a region of fixed
celestial coordinates. In some cases, instrumental constraints (i.e. the 
ability to rotate the telescope or detector array relative to the optical 
boresight)
may not allow alignment between the detector array and the desired survey
grid. This will result in coverage holes in the surveyed region, unless 
the grid spacing is reduced enough to prevent holes regardless of the 
array orientation. If a square grid with a spacing of $\Delta X = X/\sqrt{2}$
is used then coverage holes are prevented for any possible orientation
of the arrays. This is illustrated by plots in the first two rows of 
Figure 9, which shows the array positions for $4\times 4$ $M=1$ survey
(without dithering). With a deep survey strategy, avoidance of holes in the 
$M=1$ case will prevent holes at any depth $M$, but for the shallow survey 
strategy additional overlap may need to be built into the survey grid
to prevent holes as discussed above. Decreasing the survey grid by a 
factor of $\sqrt{2}$ in each dimension results in a grid that covers only
half the area that could be covered if the detectors and grid are aligned.
This efficiency can be increased if the survey is set up on a triangular grid 
rather than a square grid. If alternate rows of the survey grid are 
staggered by $X/2$ (middle row of Fig. 9) and the vertical spacing 
of the grid is reduced by a factor of $\sqrt{3}/2$, then holes are prevented
as long as the array orientation remains fixed throughout the survey (4th 
row of Fig. 9). The area covered by this triangular grid will be $\sim~87\%$
of the maximum possible area, rather than $50\%$ for the square grid required
to prevent holes. If the array orientation is not fixed throughout the survey 
(last column of Fig. 9) then the triangular grid must be reduced by an 
additional factor of $\sqrt{3}/2$ in both dimensions. This results in a 
$\sim65\%$ efficiency for the triangular grid versus $50\%$ for the square
grid, which requires no further reduction. The FOM of a survey on a triangular
grid is similar to that of a survey on a square grid with an equivalent amount 
of overlap.
\begin{figure}[t]
\epsscale{0.75}
\plotone{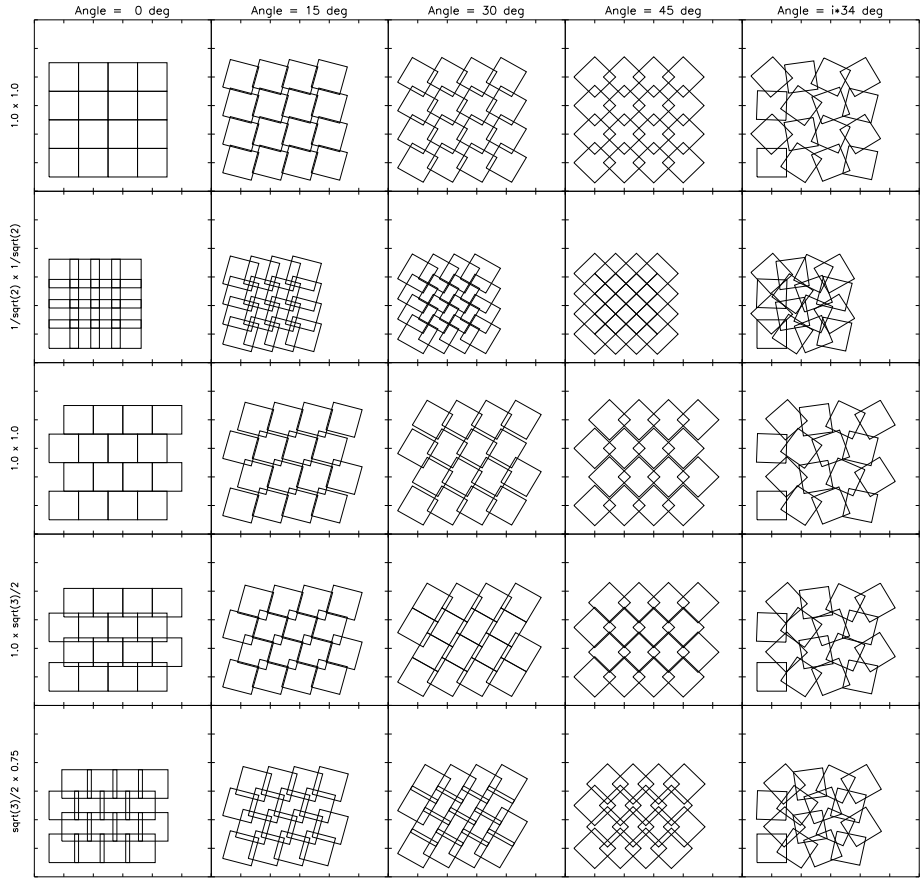}
\caption[]{Examples of $4\times 4$ $M=1$ surveys on square 
(1st and 2nd rows), staggered (middle row), and triangular grids (4th and 5th
rows) for various angles between the detector array and the grid orientation.
In the last column the array orientation was rotated by $34\arcdeg$ at each
successive survey grid point. The squares indicate the outline of the entire 
array as pointed at each survey grid point.}
\end{figure}

\section{Other Miscellaneous Details}
The most flexible implementation of the dithering strategies presented here
would be to have the dither steps be determined algorithmically from 
a small set of user-supplied parameters. For example, an observer could
select: a type of dither pattern (e.g. Reuleaux triangle or random),
a pattern depth $M_{pattern}$, and a scaling factor to control the overall size
of the pattern. From this information, the telescope control software could
calculate and execute the desired dither pattern. For the shallow survey 
strategy presented above, the observer would also need to supply:
the survey depth, $M_{survey} < M_{pattern}$, and perhaps an index to track 
which grid point of the survey is being considered (software might handle 
this automatically). 

Sometimes design or operational constraints require that the 
dither patterns reside in
a set of pre-calculated look-up tables. In this case (which has applied
to both WIRE and IRAC) the observer's ability to set the dither pattern is 
more limited. However, some of the limitations of using dither tables can be 
mitigated if the observer is not forced to use dither steps from 
the tables in a strictly sequential fashion. 
For example, one dither table might contain an $M=72$ 
Reuleaux triangle dither pattern calculated on a scale to produce the
optimum FOM. If the observer is allowed to set the increment, $\Delta i$, 
used in stepping through this dither table, then by selecting 
$\Delta i = 3$ or $\Delta i = 4$, then dither patterns of $M=24$ or $M=18$ 
can be generated. Allowing non-integer increments (subsequently rounded) 
would enable the selection of a dither pattern of any depth $M\leq 72$. 
This adjustment of the 
increment is most clearly useful for very symmetric dither patterns such as
the Reuleaux triangle pattern. For a dither table containing a 
random pattern, non-sequential access to the table can have other uses.
First, in applying the shallow survey strategy, a random dither table
of length $M_{pattern}$ could be used to sequentially generate 
$M_{survey} < M_{pattern}$ dithers at each successive survey grid point. 
Selection of dither steps would wrap around to the beginning of the table
once the end of the table is reached. For example a dither table of
$M_{pattern} = 100$ could be used sequentially to generate 20 different 
patterns for an $M=5$ shallow survey. Even better would be to have a
table with $M_{pattern}$ a prime number, e.g. 101. Then, wrapping the 
table allows the sequential generation of $M_{pattern}$ different dither
patterns for any $M_{survey}$, though some of these dither patterns 
will differ from others by only one step. Additional random patterns can
be generated by setting different increments for stepping through the table. 
Enabling specification of the starting point in the dither table would
additionally allow the observer to pick up the random dither pattern sequence
at various (or the same) positions as desired. These capabilities would
enable an observer to exploit the large number of {\it combinations} of dither
steps available in a finite length dither table, in efforts to maximize 
the FOM. 
Use of a fixed dither table can also be made less restrictive if a scaling 
factor can be applied to the dither pattern size.
A free scaling factor provides an additional means of adjusting the 
pattern size as desired to meet coverage or FOM goals. 

For the cases presented in this paper, we have assumed that the orientation
of the detector array remains fixed throughout the execution of the dither
pattern and any larger survey (except for the last column of Fig. 9).
However, rotation of the detector array relative to the dither pattern,
either within a single pointing, or at different pointings in a deep survey,
is an effective way of establishing combinations of direct pixel-to-pixel 
correlations that cannot be obtained using purely translational dither steps.
Inclusion of rotation of the detector can lead to further improvements
in the FOM of a given dither pattern or survey. In the extreme, a dither
pattern could even be made entirely out of rotational rather than 
translational dither steps. However, without an orthogonal ``radial'' dither
step, rotation alone is similar to dithering with steps in the $x$-direction 
but not the $y$-direction. The ability to implement rotations of the 
detector will be allowed or limited by the design and operating 
constraints of the telescope and instruments being used.

Bright sources can often saturate detectors and cause residual time-dependent 
variations in detector properties. For observations of a field containing 
a bright source, use of a random dither pattern may lead to 
streaking as the source is trailed back and forth across the detector 
array between dithers. In contrast the use of a basically hollow or 
circular dither pattern such as the Reuleaux triangle pattern, will only 
trail the source through a short well-defined pattern, which will lie 
toward the outer edge of the detector if the source position is centered
in the dither pattern. If the pattern scale of the dither pattern is increased,
the trail of the source can be pushed to or off the edges of the detector,
though the $FOM$ will suffer if the pattern scale is greatly increased.
In other words, a hollow dither pattern with a large scale could be used to 
obtain a series of images looking around but not at a bright source.

Dithering may be performed by repointing the telescope, or by repositioning
the instrument in the focal plane, for example through the use of a 
tilting optics as in the 2MASS (Kleinmann 1992) or 
{\it SIRTF} MIPS (Heim, et al. 1998) instruments.
Calculation of the $FOM$ of the dither pattern will be independent of the 
technique used. The self-calibration procedure, however, may be affected by
effective instrumental changes if it is repositioned in the focal plane.
The alternative repointing of the telescope can be much more time consuming 
and may limit the use of large $M$ dither patterns.

The combined use of two or more non-contiguous fields is transparent to the
self-calibration procedure. If the same dither pattern is used on each of
the separate fields, the resulting $FOM$ will be the same as that for a single
field. The $FOM$ would be improved for the combined data set if the dither
pattern is different for each of the subset. The $FOM$ for data set 
of non-contiguous regions is thus similar to that obtained using the same
dither strategy in a contiguous survey, except there is a small loss in the
$FOM$ because of the lack of overlap between adjacent regions.

Another means of minimizing coverage holes when using a shallow survey strategy
is to oversample the depth of the survey. For example, performing the 
shallow survey at a depth of $M=4$ when $M=3$ is the intended goal 
will result in fewer holes at a depth of 3 for a fixed grid spacing, and in 
a better FOM for the overall survey. However, the cost in time of the 
additional exposures may be prohibitive.

The FOM as calculated here only depends upon the offsets of the dither pattern
rounded to the nearest whole pixel. This means that any desired combination
of fractional pixel offsets to facilitate subpixel image reconstruction may
be added to the dither patterns without affecting the various aspects discussed 
in this paper. If using dither tables, one could have separate tables for the 
large scale and the fractional pixel dithers, with the actual dithers made 
by adding selected entries from the two tables. This could allow 
simultaneous and independent implementation of large-scale and subpixel 
dithering strategies.
Only subpixel image reconstruction that demands exclusively
small ($\sim1$ pixel) dithering would be incompatible with the dithering
strategies presented here.

\section{Conclusion}
We have shown that proper selection of observing strategies can 
dramatically improve the quality of self-calibration of imaging detectors.
We have established a figure of merit (Eq. 6) for quantitatively ranking 
different dither patterns, and have identified several patterns
that enable good self-calibration of a detector on
all spatial scales. The layouts of radio interferometers 
correspond to good dither patterns. Both the highly ordered Reuleaux triangle 
pattern and the unstructured random pattern provide good FOM with moderate
or deep observations. This indicates that good patterns must sample a range 
of spatial scales without redundancy, and if this condition is met, 
then secondary characteristics of the patterns or instrument constraints
may determine the actual choice of the dither pattern. Any dither pattern 
must contain steps as large as half the size of the detector array if large 
scale correlations are to be effectively encoded in the dithered data set.
Deep surveys can take advantage of the use of a single good dither pattern.
Shallow surveys can obtain good FOM by altering the dithers used at each of 
the survey grid points. Using a fixed pattern 
throughout a shallow survey makes it
difficult or impossible to apply a self-calibration procedure to the 
resulting data sets. The use of triangular instead of square survey grids
can be more efficient in executing complete-coverage surveys when the array 
orientation cannot be set to match the survey grid. Good dither patterns 
and survey strategies can be devised even in some seemingly restricted 
situations. The ultimate importance of dithering and a good FOM will depend
on the nature of the instrument and the data and on the scientific goals.
For many goals, obtaining a larger quantity of data may not be an adequate
substitute for obtaining data with a good FOM. 

\acknowledgements
We thank D. Shupe and the WIRE team for supplying simulated data using several
different dither patterns. W. Reach and members of the SSC and 
IRAC instrument teams were helpful in providing useful ideas and criticism 
throughout the development of this work. J. Gardner, J. Mather, and the 
anonymous referee provided very helpful criticism of the manuscript.

\begin{deluxetable}{ll}
\footnotesize
\tablewidth{0pt}
\tablecaption{Comparison of Data Reduction Procedures}
\tablehead{
\colhead{Flat Sky Technique} &
\colhead{Least Squares Solution} 
}
\startdata
assumes sky is flat ($S^0$) & 
solves for real sky ($S^{\alpha}$); will find $S^0$ if warranted\\
 & \\
requires dark frames &
no dark frames needed, but they are useful if available\\
 & \\
may take time for chopping to nearby &
no chopping needed\\
flat field region (if such exists) &
\\
 & \\
confused fields can ruin the solution & 
confused fields can improve the solution\\
 & \\
may remove flat emission components of the astronomical sky &
preserves full sky intensity\\
(e.g. zodiacal emission, nebular emission, cosmic backgrounds) &
\\
 & \\
requires Monte Carlo or ad hoc assessment of uncertainties & 
can accurately and analytically track\\
or unbiased source removal & 
uncertainties and correlations\\
 & \\
observations of $S^0$ by all pixels automatically tightly & 
dithering must establish tight correlations between \\
correlate all detector pixels on all spatial scales &
all detector pixels\\ 
 & \\
computationally simple & 
can be simplified to produce the flat sky result\\
 & \\
  & can be used in non-astronomical applications\\
\enddata
\end{deluxetable}

\begin{deluxetable}{cccccccccccc}
\tiny
\tablewidth{0pt}
\tablecaption{Figures of Merit for Single Pointings}
\tablehead{
 & \multicolumn{8}{c}{$256\times 256$ arrays} & & \multicolumn{2}{c}{$32\times 32$ arrays} \\
\cline{2-9}\cline{11-12}
\colhead{$M_{pattern}$} &
\colhead{Reuleaux} &
\colhead{Random} & 
\colhead{VLA} &
\colhead{Geometric} &
\colhead{DASI} &
\colhead{WIRE} &
\colhead{NICMOS HDF-S} &
\colhead{2MASS} &
\colhead{} &
\colhead{Reuleaux} &
\colhead{Grid} 
}
\startdata
   6 & 0.002   & \nodata & 0.020\tablenotemark{e}   & 0.016   & \nodata & \nodata & \nodata & 0.002   & & \nodata & \nodata \\
   8 & \nodata & \nodata & \nodata & 0.058   & \nodata & \nodata & \nodata & \nodata & & \nodata & \nodata \\
   9 & 0.003\tablenotemark{a}   & \nodata & 0.059\tablenotemark{f}   & \nodata & \nodata & \nodata & \nodata & \nodata & & 0.217\tablenotemark{k}   & \nodata \\
  10 & \nodata & \nodata & \nodata & 0.094   & \nodata & \nodata & \nodata & \nodata & & \nodata & \nodata \\
  12 & \nodata & \nodata & \nodata & 0.112   & \nodata & \nodata & \nodata & \nodata & & \nodata & \nodata \\
  13 & \nodata & \nodata & \nodata & \nodata & 0.100   & \nodata & \nodata & \nodata & & \nodata & \nodata \\
  14 & \nodata & \nodata & \nodata & 0.139   & \nodata & \nodata & \nodata & \nodata & & \nodata & \nodata \\
  15 & 0.117\tablenotemark{b}   & \nodata & 0.128\tablenotemark{g}   & \nodata & \nodata & \nodata & \nodata & \nodata & & \nodata & \nodata \\
  16 & \nodata & 0.162   & \nodata & 0.152   & \nodata & \nodata & \nodata & \nodata & & \nodata & \nodata \\
  18 & 0.153\tablenotemark{c}   & 0.181   & 0.166   & 0.133   & \nodata & \nodata & \nodata & \nodata & & \nodata & \nodata \\
  24 & \nodata & \nodata & \nodata & 0.168   & \nodata & \nodata & \nodata & \nodata & & \nodata & \nodata \\
  27 & 0.265   & \nodata & 0.240\tablenotemark{h}   & \nodata & \nodata & \nodata & \nodata & \nodata & & \nodata & \nodata \\
  32 & \nodata & 0.286   & \nodata & 0.193   & \nodata & \nodata & \nodata & \nodata & & \nodata & \nodata \\
  39 & 0.307   & \nodata & 0.282   & \nodata & \nodata & \nodata & \nodata & \nodata & & \nodata & \nodata \\
  40 & \nodata & \nodata & \nodata & 0.193   & \nodata & 0.228\tablenotemark{i}   & \nodata & \nodata & & \nodata & \nodata \\
  60 & 0.323   & 0.318   & 0.303   & 0.204   & \nodata & \nodata & \nodata & \nodata & & \nodata & \nodata \\
  90 & 0.341   & 0.335   & 0.316   & 0.208   & \nodata & \nodata & \nodata & \nodata & & \nodata & \nodata \\
 120 & 0.361   & 0.351   & 0.329   & 0.208   & \nodata & 0.225\tablenotemark{j}   & \nodata & \nodata & & \nodata & \nodata \\
 142 & \nodata & \nodata & \nodata & \nodata & \nodata & \nodata & 0.131   & \nodata & & \nodata & \nodata \\
 150 & 0.387   & 0.365   & 0.347   & \nodata & \nodata & \nodata & \nodata & \nodata & & \nodata & \nodata \\
 180 & 0.416   & 0.378   & 0.366   & \nodata & \nodata & \nodata & \nodata & \nodata & & \nodata & \nodata \\
 210 & 0.448   & 0.415   & 0.392   & \nodata & \nodata & \nodata & \nodata & \nodata & & \nodata & \nodata \\
 240 & 0.485\tablenotemark{d}   & 0.437   & 0.419   & 0.212   & \nodata & \nodata & \nodata & \nodata & & \nodata & \nodata \\
 300 & 0.526   & 0.439   & 0.469   & \nodata & \nodata & \nodata & \nodata & \nodata & & \nodata & \nodata \\
1024 & \nodata & \nodata & \nodata & \nodata & \nodata & \nodata & \nodata & \nodata & & \nodata & 0.783   \\
4096 & \nodata & \nodata & \nodata & \nodata & \nodata & \nodata & \nodata & \nodata & & \nodata & 0.889   \\
\enddata
\tablecomments{Standard pattern sizes are: Reuleaux width = 128 pixel, Random 3 
$\sigma$ = 128 pixel, VLA $r_{max}$ = 125.7 pixel, Geometric is scaled to the 
256 pixel array size, DASI $r_{max}$ = 82.5 pixel, WIRE medium and deep surveys
scaled by a factor of 2, NICMOS HDF-S all camera 3 F110W data, $32 \times 32$
Reuleaux width = 12 pixel, Grid spacing = 1 pixel}
\tablenotetext{a}{0.037 if 3\% random variations added}
\tablenotetext{b}{0.146 if 3\% random variations added}
\tablenotetext{c}{0.171 if 3\% random variations added}
\tablenotetext{d}{0.396 for width = 110 pixel, 0.513 for width = 144 pixel}
\tablenotetext{e}{$r_{max}$ = 16 pixel}
\tablenotetext{f}{$r_{max}$ = 32 pixel}
\tablenotetext{g}{$r_{max}$ = 125 pixel}
\tablenotetext{h}{0.176 for $r_{max}$ = 16 pixel, 0.199 for 
$r_{max}$ = 16 pixel}
\tablenotetext{i}{0.065 for the unscaled pattern}
\tablenotetext{j}{0.062 for the unscaled pattern}
\tablenotetext{k}{0.162 for width = 8 pixel}
\end{deluxetable}

\newpage
\begin{deluxetable}{ccccccccccccccccc}
\tiny
\tablewidth{0pt}
\tablecaption{Figures of Merit for Deep Surveys}
\tablehead{
\colhead{Survey} &
\colhead{Spacing} &
\colhead{} &
\multicolumn{2}{c}{Reuleaux} &
\colhead{} &
\multicolumn{3}{c}{Random} &
\colhead{} &
\multicolumn{3}{c}{VLA} &
\colhead{} &
\multicolumn{3}{c}{Geometric} \\
\cline{4-5}\cline{7-9}\cline{11-13}\cline{15-17}
\colhead{Size} &
\colhead{(pixels)} &
\colhead{} & 
\colhead{$M = 15$} &
\colhead{39} &
\colhead{} &
\colhead{$M = 6$} &
\colhead{16} &
\colhead{40} &
\colhead{} &
\colhead{$M = 6$} &
\colhead{15} &
\colhead{39} &
\colhead{} &
\colhead{$M = 6$} &
\colhead{16} &
\colhead{40}
}
\startdata
$2\times 2$ & 181 & & 0.194   & 0.318   & & 0.028   & 0.209   & 0.318   & &
            0.028   & 0.156   & 0.310   & & 0.027   & 0.183   & 0.221   \\
$2\times 2$ & 218 & & 0.173   & 0.314   & & 0.016   & 0.191   & 0.313   & &
            0.022   & 0.168   & 0.303   & & 0.022   & 0.174   & 0.216   \\
$2\times 2$ & 256 & & 0.166   & 0.311   & & 0.012   & 0.172   & 0.309   & &
            0.020   & 0.153   & 0.298   & & 0.020   & 0.165   & 0.211   \\
$3\times 3$ & 181 & & 0.198   & \nodata & & 0.020   & 0.213   & \nodata & &
            0.028   & 0.160   & \nodata & & 0.027   & 0.187   & \nodata \\
$3\times 3$ & 218 & & 0.182   & \nodata & & 0.017   & 0.194   & \nodata & &
            0.022   & 0.172   & \nodata & & 0.022   & 0.177   & \nodata \\
$3\times 3$ & 256 & & 0.170   & \nodata & & 0.016   & 0.178   & \nodata & &
            0.020   & 0.158   & \nodata & & 0.020   & 0.168   & \nodata \\
$4\times 4$ & 181 & & \nodata & \nodata & & 0.022   & \nodata & \nodata & &
            0.028   & \nodata & \nodata & & 0.027   & \nodata & \nodata \\
\enddata
\end{deluxetable}

\begin{deluxetable}{cccccccccccccccc}
\scriptsize
\tablewidth{0pt}
\tablecaption{Figures of Merit for $4\times 4$ Shallow Surveys}
\tablehead{
\colhead{Survey} &
\colhead{} &
\multicolumn{2}{c}{Reuleaux} &
\colhead{} &
\multicolumn{2}{c}{Random} &
\colhead{} &
\multicolumn{2}{c}{VLA} &
\colhead{} &
\multicolumn{2}{c}{Geometric} &
\colhead{} &
\multicolumn{2}{c}{Grid} \\
\cline{3-4}\cline{6-7}\cline{9-10}\cline{12-13}\cline{15-16}
\colhead{Depth} &
\colhead{} & 
\colhead{$M = 16$} &
\colhead{32} &
\colhead{} &
\colhead{$M = 16$} &
\colhead{32} &
\colhead{} &
\colhead{$M = 16$} &
\colhead{32} &
\colhead{} &
\colhead{$M = 16$} &
\colhead{32} &
\colhead{} &
\colhead{$M = 2$} &
\colhead{3}
}
\startdata
2 & & 0.088 & 0.117   & & 0.086   & 0.120   & & 0.090 & 0.120 & & 
      0.079 & 0.077   & & 0.001   & \nodata \\
3 & & 0.113 & 0.175   & & 0.115   & 0.177   & & 0.109 & 0.168 & & 
      0.110 & 0.135   & & \nodata & 0.002 \\
3\tablenotemark{a} & & 0.182 & \nodata & & 0.165   & \nodata & & 0.167 & \nodata & & 
      0.136 & \nodata & & \nodata & \nodata \\
\enddata
\tablecomments{All surveys used 181 pixel grid spacing.}
\tablenotetext{a}{Random rather than sequential selections 
from the dither patterns.}
\end{deluxetable}

\end{document}